\documentclass[10pt,conference]{IEEEtran}
\usepackage{tikz}
\usepackage{pifont}
\usepackage{dashrule}
\usepackage{listings}
\usepackage{caption}
\usepackage{tcolorbox}
\usepackage{xcolor}
\usepackage{hyperref}

\usepackage{booktabs}
\usepackage{svg}
\usepackage{multirow}
\usepackage[numbers]{natbib}

\usepackage{tikz}
\usepackage{xcolor}
\usepackage{amsmath}
\usetikzlibrary{shapes,arrows,positioning,calc,fit,backgrounds}
\usepackage{cleveref}

\usepackage{listings}

\newcommand{\header}[1]{\smallskip\noindent\textbf{#1}}

\newcommand{\rqi}{To what extent does GPT-4o mini log in the same position as human written logs?} 
\newcommand{\rqii}{How does GPT-4o mini perform at logging ML applications compared to human developers?} 
\newcommand{\rqiii}{What are the challenges of using GPT-4o mini for automated log generation?} 

\begin{document}

\title{Automated File-Level Logging Generation for Machine Learning Applications using LLMs: A Case Study using GPT-4o Mini}

\author{
\IEEEauthorblockN{Mayra Sofia Ruiz Rodriguez}
\IEEEauthorblockA{\textit{Concordia University} \\
Montreal, Canada \\
m\_ruizro@live.concordia.ca}
\and
\IEEEauthorblockN{SayedHassan Khatoonabadi}
\IEEEauthorblockA{\textit{Concordia University} \\
Montreal, Canada \\
sayedhassan.khatoonabadi@concordia.ca}
\and
\IEEEauthorblockN{Emad Shihab}
\IEEEauthorblockA{\textit{Concordia University} \\
Montreal, Canada \\
emad.shihab@concordia.ca}
}

\maketitle

\begin{abstract}
    Logging is essential in software development, helping developers monitor system behavior and aiding in debugging applications. Given the ability of large language models (LLMs) to generate natural language and code, researchers are exploring their potential to generate log statements. However, prior work focuses on evaluating logs introduced in code functions, leaving file-level log generation underexplored---especially in machine learning (ML) applications, where comprehensive logging can enhance reliability. In this study, we evaluate the capacity of GPT-4o mini as a case study to generate log statements for ML projects at file level. We gathered a set of 171 ML repositories containing 4,073 Python files with at least one log statement. We identified and removed the original logs from the files, prompted the LLM to generate logs for them, and evaluated both the position of the logs and log level, variables, and text quality of the generated logs compared to human-written logs. In addition, we manually analyzed a representative sample of generated logs to identify common patterns and challenges. We find that the LLM introduces logs in the same place as humans in 63.91\% of cases, but at the cost of a high overlogging rate of 82.66\%.  Furthermore, our manual analysis reveals challenges for file-level logging, which shows overlogging at the beginning or end of a function, difficulty logging within large code blocks, and misalignment with project-specific logging conventions. While the LLM shows promise for generating logs for complete files, these limitations remain to be addressed for practical implementation.
\end{abstract}

\begin{IEEEkeywords}
    Logging practice, ML-based applications, large language models, mining software repositories
\end{IEEEkeywords}

\section{Introduction}
Logging plays a critical role in modern software development, enabling monitoring in production environments \cite{oliner2012advances}, failure diagnosis \cite{yuan2012characterizing, yuan2010sherlog}, and facilitating performance analysis \cite{xu2009detecting}. For effective logging, developers need to decide what-to-log and where-to-log. What-to-log refers to choosing the appropriate log level (e.g., warning, error), text, and variables, while where-to-log concerns the location in the code where the log should be introduced. With the advent of large language models (LLMs), researchers have seen promising results in various software development tasks, including code generation \cite{li2025structured}, testing \cite{kang2023large}, and resolving bugs \cite{xia2023automated}.

Recently, researchers have worked on the non-trivial task of generating complete log statements using LLMs. Mastropaolo et al. \cite{mastropaolo2022lance} introduced LANCE and later LEONID \cite{mastropaolo2024deep}, both models trained on millions of methods to generate log statements at the method level. However, the fine-tuning process for these models requires significant computational resources. To reduce training demands, Xu et al. \cite{xu2024unilog} proposed a framework that uses an in-context learning prompt paradigm with a few carefully selected examples to guide the model to generate logs. Building on this direction, Li et al. \cite{li2023exploring} evaluated different LLMs' performance on log generation. Their study involves removing log statements from functions and using placeholders where logs should be introduced before prompting the LLM.

Prior studies have primarily focused on inserting a single log within a method, without providing the LLM with the full context of the file and without considering the possibility of inserting multiple logs. Furthermore, compared to traditional software, machine learning (ML) applications face distinct characteristics: there is an inherent dependency on data, and their non-deterministic behavior makes it difficult to debug. ML applications show lower log statements than traditional software, averaging one log statement per 1150 lines of code in ML applications using Python \cite{foalem2024logging_ml}. As a result, using LLMs for this purpose is not systematically explored.

In this study, we aim to investigate how GPT-4o mini performs in the task of logging complete ML files, without specifying any placeholder in the code. To this end, we first curate a list of 371 ML projects on GitHub. We extract all Python files from each project and determine which Python files have at least one log, reducing our dataset to 171 projects. We collect a dataset consisting of 4,073 Python files, then we remove the logs in the files and prompt GPT-4o mini to generate logs for these files. Based on this dataset, we aim to answer the following three research questions: \\

\header{RQ1: {\rqi}}
 We aim to understand human and LLM logging patterns because, while logging is essential for debugging, LLMs' tendency toward verbose outputs \cite{saito2023verbosity} might result in excessive logging. We compared the position of logs for both human-written and GPT-4o mini generated logs. We find that the LLM prioritizes comprehensive logging coverage, as shown by its low underlogging rate (4.75\%) and moderately high coverage (63.91\%). However, that results in generating a large number of logs: 5.15 times more than those written by humans. Even in cases where both humans and the LLM logged, the LLM tends to include more logs overall, with a rate of 68.03\%. \\
 
\header{RQ2: {\rqii}}
 We aim to understand the quality of LLM-generated logs compared to human-written logs. To this end, we analyze the generated logs in terms of ingredients: level, variables, and text; and compare them to the corresponding human-generated log. We find that GPT-4o mini shows moderate quality but significant limitations when generating logs compared to human developers. While it matches the log levels with 59.19\% exact match and 84.34\% average ordinal distance, it identifies only 40.58\% of variables that human logged. Its log texts are similar in meaning to human logs (ROUGE-L: 0.316), but uses different vocabulary and phrasing (BLEU-4: 0.050), and require substantial editing to match human logs (Levenshtein: 0.735). \\
    
\header{RQ3: {\rqiii}} 
 We aim to identify GPT-4o mini's limitations in file-level logging. For this purpose, we categorize all developer-LLM pairs of logs into four categories: overlogging, underlogging, different levels, and different variables. We get a stratified sample of 384 pairs of logs, and manually analyze the surrounding code in which they were introduced. We find that GPT4-o mini's main challenge is overlogging (85.8\%), with most logs introduced at the start or end of a function. Moreover, the LLM underlogs (4.7\%), particularly in large code blocks. In cases with different log level (5.3\%), the LLM does not align with project-specific logging conventions. Finally, in case of different variables captured between humans and the LLM (4.3\%), the LLM sometimes misses important variables that are present in the GT, despite the larger available context.

\medskip

Based on our findings, we discuss key considerations for using GPT-4o mini in logging complete ML files. While LLMs can generate useful log statements, they often overlog, potentially cluttering code and reducing its effectiveness for debugging. Developers should critically review LLM-generated logs to ensure relevance and avoid excessive logging. Additionally, repository-specific logging configurations, such as custom log levels, are generally overlooked but crucial for log generation in real-world projects. Our results suggest that future tools might benefit from this information to generate higher-quality logs that aligns better with previous developers' preferences.

\header{Our Contributions.}
In summary, we make the following contributions in this paper:

\begin{itemize}
    \item We provide empirical evidence on the effectiveness of GPT-4o mini for automated file-logging generation.
    \item We discuss the challenges of using GPT-4o mini for file-level log generation based on our findings.
    \item To promote the reproducibility of our study and facilitate future research on this topic, we publicly share our scripts and dataset.\footnote{https://zenodo.org/records/15558239}
\end{itemize}

\header{Paper Organization.}
The rest of this paper is organized as follows. \Cref{sec:study_design} describes our study design. \Cref{sec:results} presents the findings of our three research questions. \Cref{sec:implications} discusses the implications of our findings. \Cref{sec:limitations} outlines the threats to the validity of our study. \Cref{sec:related_work} discusses related work. Finally, \Cref{sec:conclusion} concludes this paper.

\section{Study Design}
\label{sec:study_design}
\begin{figure*}
    \centering
\includegraphics[width=\linewidth]{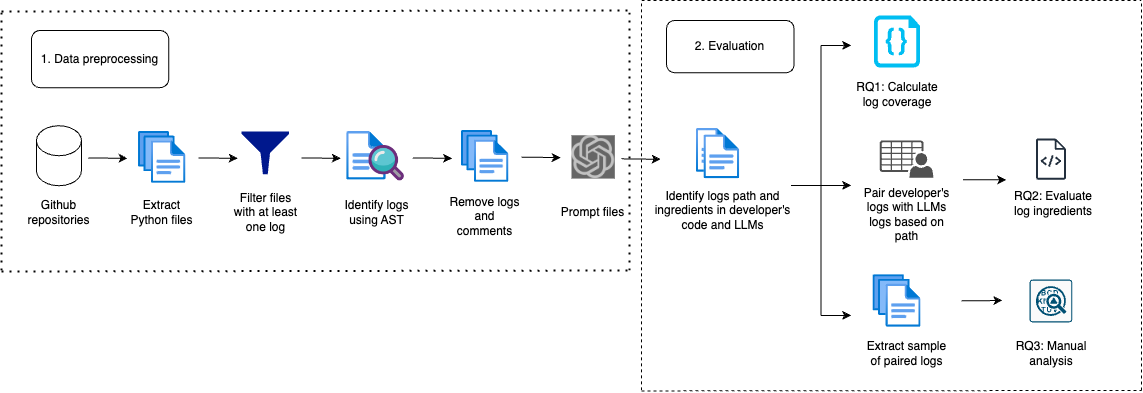}
    \caption{Research workflow of this study}
    \label{fig:research-workflow}
\end{figure*}

This study aims to evaluate the effectiveness of GPT-4o generated logs for ML applications at the file level. \Cref{fig:research-workflow} provides an overview of our approach, and we explain the details in the following sections.

\subsection{Dataset}

\begin{table}
\centering
\caption{Descriptive Statistics of the Selected Projects}
\begin{tabular}{lrrrr}
\toprule
 & \textbf{Mean} & \textbf{Median} & \textbf{Min} & \textbf{Max} \\
\midrule
\# Stars        & 5,393 & 886 & 54 & 133,592 \\
\# Commits      & 4,000 & 1,159  & 12  & 79,758  \\
\# Contributors & 78   & 25   & 3   & 455   \\
\# Days since last push & 78   & 27   & 0   & 364   \\
Age (years)     & 7.5    & 7    & 3.2   & 15.1    \\
\bottomrule
\label{tab:project-statistics}
\end{tabular}
\end{table}

\header{Project Selection.} To analyze logging in ML projects, we need a comprehensive collection of ML repositories. For this purpose, we use the dataset originally developed by Gonzalez et al. \cite{gonzalez2020ml_universe} and later revised by Rzig et al. \cite{rzig2022characterizing}. The dataset was built by identifying GitHub topic labels related to ML and collecting matching projects via the GitHub API. Rzig et al. further improved the dataset through manual analysis, removing non-ML and trivial projects, and adding an additional layer of human verification. The final dataset consists of 4,031 ML projects comprising both ML frameworks like TensorFlow and ML applications like Faceswap.

To select projects for our study, we use the GitHub API to mine the metadata of each project from the above dataset as of October 2024 and filter our dataset by choosing repositories that meet the following criteria:

\begin{itemize}
  \item We chose Python as the target language for our study, due to its popularity in ML tools and applications \cite{gonzalez2020ml_universe}.
  \item We focus on projects with at least 50 number of stars and at least three contributors to filter out toy projects.
  \item We select projects that are active by filtering the projects that have their last push operation within the last 365 days. This ensures that the selected projects capture the most current logging activity from developers.
\end{itemize}

This filtering process concludes with a dataset of 371 ML projects. With this dataset, we create a shallow clone of each repository to extract its Python files, resulting in a collection of 105,332 Python files. We then apply an additional filter to select only Python files using the Python \textit{logging} library and containing at least one log statement, which results in a total of 4,073 files across 171 projects. \Cref{tab:project-statistics} contains descriptive statistics of the selected projects. From these selected files, we removed all existing log statements and comments to prepare them for our analysis.

\subsection{AST Visitor} We utilize the Abstract Syntax Tree (AST) Python module \cite{ast_docs} to precisely locate logging positions within the source code. The AST provides a tree representation of the abstract syntax tree of source code, where each node corresponds to different syntactic elements such as class definitions, function declarations, control flow constructs (i.e. \textit{if}, and \textit{while} statements), among other node types defined in the AST module. For each file, we generate an AST and utilize the visitor pattern to traverse and examine each node within the tree.

\header{Define path.} In this study, we define "path" as the sequence of consecutive code blocks where a log is positioned. While traversing the AST, we maintain a record of the visited code blocks (i.e., classes, functions, for statements). First, we initialize the path as "global" to refer to the log statements located outside functions and classes. As we navigate through each code block type, we employ a stack data structure to track the nodes we enter. For each node, we examine whether it contains a log statement. When we exit the node, we remove the code block name from our stack, thus updating the current path definition.

We address cases involving multiple code-blocks with identical names, such as multiple \textit{if} statements at the same hierarchical level within a function. For such cases, we track the nodes: \textit{if, for, else, while, try, except,} and \textit{with}. When the AST visitor encounters these nodes, we extend our path definition. Instead of simply appending the node name, we also append a sequential number to distinguish between repeated instances of the same node type within the same code block. In \Cref{fig:code_example}, the log statement on line 3 will have the path "global/Analysis/\_\_init\_\_". The log statement on line 6 will have the path "global/Analysis/\_\_init\_\_/if1", while the log statement on line 10 will have the path "global/Analysis/\_\_init\_\_/if2".

\begin{figure}
\centering
\begin{minipage}{0.43\textwidth}
\begin{lstlisting}
class Analysis():
    def __init__(self, data, message):
        logger.info("Initializing")

        if data is None:
            logger.debug("Data not yet available")
        else:
            logger.debug("Data available")
        if message is None:
            logger.debug("No message received")
\end{lstlisting}
\end{minipage}
\caption{Example Python code}
\label{fig:code_example}
\end{figure}

Nodes like \textit{else} are inherently dependent on their corresponding \textit{if} statements. In our approach, we prioritize the order of node entry and deliberately do not associate \textit{if}/\textit{else} or \textit{try}/\textit{except} pairs with a shared numerical identifier. For example, in line 8 of \Cref{fig:code_example}, since we visit the \textit{if} node, then exit and enter the \textit{else} node, we report the path as "global/Analysis/\_\_init\_\_/else1".

Our approach allows us to assign each log a position based on its location in the AST, rather than its line number in the source code. Finally, we identify log statements in our dataset, we drop them from the files, and we prompt GPT-4o mini to generate the logs for these files.

\subsection{Prompt Crafting} 
\begin{figure}
    \centering
    \begin{tikzpicture}
        \draw[black, thick] (-3.75,-4.25) rectangle (3.75,4.25);
        \node[align=left, text width=6.5cm, font=\small] at (-0.25,0) {
        \begin{itemize}
            \item[\ding{182}] You are an expert machine learning developer. You will receive a Python file. Follow these instructions: \\
            \hdashrule{\linewidth}{1pt}{2pt} 
            \item[\ding{183}] 1. Review the provided Python file \\
            \hdashrule{\linewidth}{1pt}{2pt}
            \item[\ding{184}] 2. Add any missing log statements using the logging library. \\
            \hdashrule{\linewidth}{1pt}{2pt}
            \item[\ding{185}] 3. Verify that each logging statement is in an appropriate position within the code. \\
            4. Check the logging level of each logging statement to ensure it aligns with its importance. \\
            5. Evaluate the quality of the log texts, ensuring they cover important details and follow best practices. \\
            \hdashrule{\linewidth}{1pt}{2pt} 
            \item[\ding{186}] 6. Return only the complete code snippet with all necessary log statements added. Do not modify the rest of the code. \\
            \hdashrule{\linewidth}{1pt}{2pt} 
            \item[\ding{187}] This is the file:
\textcolor{orange}{\$SOURCE\_CODE}\\
        \end{itemize}};
    \end{tikzpicture}
    \caption{Prompt template for automated logging}
    \label{fig:prompt}
\end{figure}
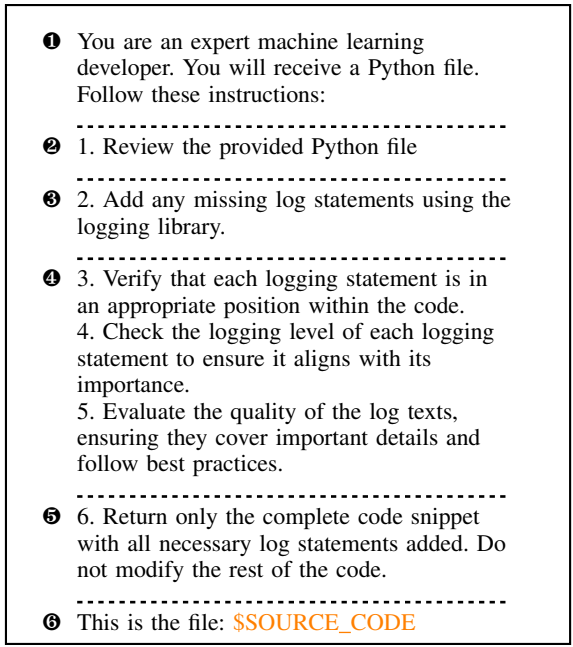

After preprocessing our files, we use the OpenAI API to prompt GPT-4o mini (\textit{gpt-4o-mini-2024-07-18}) to generate logs for each file. We use this LLM because, at the time of our study, it was a state-of-the-art model with a large context window of 128,000 tokens and a max output of 16,384 tokens, which is sufficient to process and output complete files. Following OpenAI's prompt engineering guidelines \cite{openAIPromptEngineering}, we construct a prompt illustrated in \Cref{fig:prompt}, where the "\$SOURCE\_CODE" variable is dynamically replaced with the file's content. We do not specify the exact number or position of logs, as our goal is to observe how GPT-4o mini chooses to log when asked to introduce logs for an entire file, and compare this behavior to human logs. We also set the temperature to 0 to ensure a more deterministic output \cite{li2023exploring}.

The prompt is structured into five key components:

\begin{itemize}
    \item[\ding{182}] \textbf{Persona and context.} We start the interaction by requesting the LLM to adopt the persona ``expert machine learning developer", and we inform that it will receive a Python file as input. 
    \item[\ding{183}] \textbf{Review.} We instruct the LLM to review the provided file. 
    \item[\ding{184}] \textbf{Log statement generation.} We ask the LLM to generate log statements using the \textit{logging} library. We use this library due to its prevalence in ML applications \cite{foalem2024logging_ml}. While ML-specific logging libraries such as Wandb and MLflow exist, in our dataset of 4,073 files using \textit{logging}, only 32 files used Wandb and 118 used MLFlow. Focusing on \textit{logging} allowed us to work with a larger number of examples for comparison.
    \item[\ding{185}] \textbf{Log quality instructions.} This section defines the expected log quality through precise guidelines. Each point refers to a specific log quality instruction: where-to-log and what-to-log.
    \item[\ding{186}] \textbf{Task specification.} The task the LLM is instructed to do is to return the complete file, ensuring that the only modifications are the newly introduced log statements.
\end{itemize}

\subsection{Mapping Ground Truth and LLM logs} 
Our study involves submitting complete source files to GPT-4o mini for logging, without specifying where to log or how many logs to add. The LLM is free to add as many logs as it finds necessary, sometimes exceeding the number of logs in the ground truth (GT). Due to this scenario, it can become complex to evaluate the quality of the generated logs. Moreover, in case of multiple logs in the same code block in both GT and LLM files, it is hard to know which GT log is closely related to which LLM log.

To facilitate evaluation, we establish pairings between GT and LLM-generated logs. We define four mapping possibilities:

\begin{itemize}
    \item \textbf{1:1 pairing:} A scenario where a single log exists in the same path for both GT and LLM files.
    \item \textbf{1:n pairing:} A path contains one GT log and multiple LLM-generated logs.
    \item \textbf{n:1 pairing:} A path contains multiple GT logs and a single LLM-generated log.
    \item \textbf{n:n pairing:} Multiple logs are present in both GT and LLM within the same path.
\end{itemize}

\begin{figure*}
\centering
\resizebox{0.6\textwidth}{!}{
\begin{tikzpicture}[
    node distance=1.5cm and 3cm,
    gt/.style={rectangle, draw, fill=blue!10, text width=2cm, minimum height=0.8cm, align=center},
    llm/.style={rectangle, draw, fill=green!10, text width=2cm, minimum height=0.8cm, align=center},
    arrow/.style={->, >=stealth, thick},
    box/.style={rectangle, draw, dashed, inner sep=0.4cm},
    title/.style={font=\bfseries}
]

\node[title] at (0, 0) (title1) {1:1 Pairing};
\node[gt] at (-1.5, -1) (gt1) {GT Log};
\node[llm] at (1.5, -1) (llm1) {LLM Log\\CS: \textbf{0.85}};
\draw[arrow] (gt1) -- (llm1) node[midway, above] {};

\begin{scope}[on background layer]
\node[box, fit=(gt1) (llm1) (title1)] {};
\end{scope}

\node[title] at (0, -3) (title2) {1:n Pairing};
\node[gt] at (-1.5, -4) (gt2) {GT Log};
\node[llm] at (1.5, -4) (llm2a) {LLM Log 1\\CS: \textbf{0.82}};
\node[llm] at (1.5, -5.2) (llm2b) {LLM Log 2\\CS: 0.65};
\node[llm] at (1.5, -6.4) (llm2c) {LLM Log 3\\CS: 0.47};
\draw[arrow] (gt2) -- (llm2a) node[midway, above] {};
\draw[arrow, dashed, gray] (gt2) -- (llm2b);
\draw[arrow, dashed, gray] (gt2) -- (llm2c);

\begin{scope}[on background layer]
\node[box, fit=(gt2) (llm2a) (llm2b) (llm2c) (title2)] {};
\end{scope}

\node[title] at (7, 0) (title3) {n:1 Pairing};
\node[gt] at (5.5, -1) (gt3a) {GT Log 1};
\node[gt] at (5.5, -2.2) (gt3b) {GT Log 2};
\node[gt] at (5.5, -3.4) (gt3c) {GT Log 3};
\node[llm] at (8.5, -2) (llm3) {LLM Log\\CS1: 0.71\\CS2: 0.84\\CS3: 0.45};
\draw[arrow] (gt3a) -- (llm3);
\draw[arrow] (gt3b) -- (llm3);
\draw[arrow] (gt3c) -- (llm3);

\begin{scope}[on background layer]
\node[box, fit=(gt3a) (gt3b) (gt3c) (llm3) (title3)] {};
\end{scope}

\node[title] at (7, -5) (title4) {n:n Pairing};
\node[gt] at (5.5, -6) (gt4a) {GT Log 1};
\node[gt] at (5.5, -7.6) (gt4b) {GT Log 2};
\node[llm] at (8.5, -6) (llm4a) {LLM Log 1\\CS1: \textbf{0.88}\\CS2: 0.3};
\node[llm] at (8.5, -7.6) (llm4b) {LLM Log 2\\CS1: 0.49\\CS2: \textbf{0.73}};
\node[llm] at (8.5, -9.2) (llm4c) {LLM Log 3\\CS1: 0.68\\CS2: 0.52};

\draw[arrow] (gt4a) -- (llm4a) node[midway, above] {};
\draw[arrow] (gt4b) -- (llm4b) node[midway, above] {};

\begin{scope}[on background layer]
\node[box, fit=(gt4a) (gt4b) (llm4a) (llm4b) (llm4c)(title4)] {};
\end{scope}

\node[title] at (0, -7.8) (legend) {Legend};
\node[gt, text width=3.2cm] at (0, -8.5) (legend1) {GT Log};
\node[llm, text width=3.2cm] at (0, -9.5) (legend2) {LLM-Generated Log};
\draw[arrow] (2, -8.5) -- (3, -8.5) node[midway, above] {Paired};
\draw[arrow, dashed, gray] (2, -9.5) -- (3, -9.5) node[midway, above] {Unpaired};

\end{tikzpicture}
}
\caption{Log pairing process between GT and LLM-generated logs. Each GT log is matched with its most similar LLM log based on cosine similarity (CS).}
\label{fig:pairing_examples}
\end{figure*}
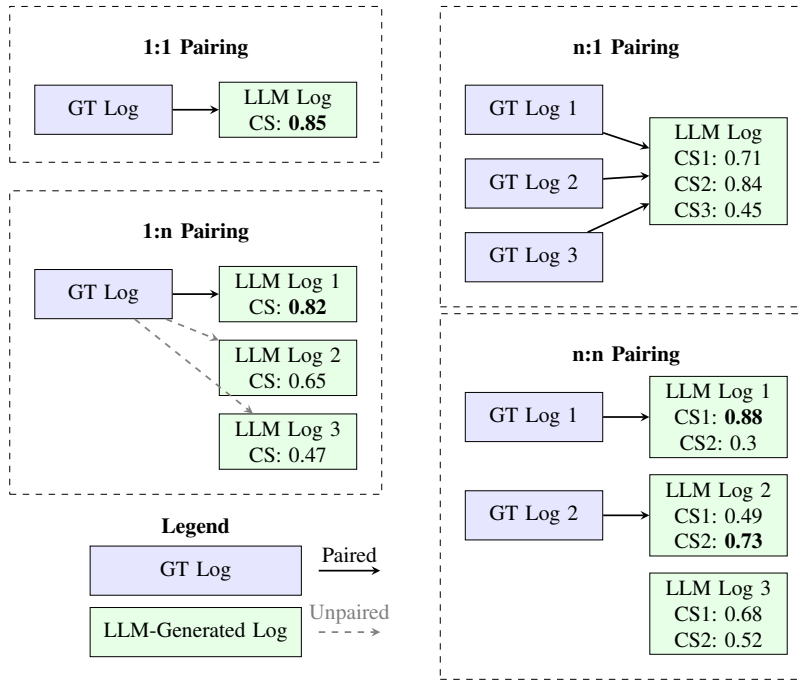

\Cref{fig:pairing_examples} illustrates these four pairing scenarios and the matching process. For the last three cases, we use cosine similarity on the whole log statement to pair each GT log with its most similar LLM-generated log. Our goal is to match each GT log to a corresponding LLM-generated log, so we can evaluate the quality of each LLM-generated log based on its code positioning. This pairing process preserves all GT logs while matching them to their semantically closest LLM-generated log. In case of 1:n pairing and n:n pairing, some LLM logs may remain unmatched if they aren't the closest match to any GT log. For instance, if one GT log corresponds to two LLM logs on the same path, we only pair it with the LLM log showing higher semantic similarity, effectively excluding the other LLM log from evaluation. To summarize, all GT logs are retained in our analysis, and each one of them will be matched to an LLM log unless the LLM produced no logs for that particular path.

\subsection{Evaluation} Using AST, we simultaneously identify log statements and their corresponding paths. We utilize two techniques to extract the log ingredients: regular expressions are used to extract variables within the log statements, while AST nodes provide information about the log level and the log text. We also remove format specifiers (i.e., \textit{\%s}, \textit{\%d}) from log texts to ensure a fair comparison. Following the path-based pairing of GT and LLM-generated logs, we proceed to evaluate the effectiveness of GPT-4o mini for generating log statements.

\section{Results}
\label{sec:results}

In this section, we discuss the results by research question. Each research question includes its motivation, the approach to answer the question, and the results.

\label{sec:rqi}
\subsection{RQ1: \rqi}

\header{Motivation.} Logging is critical for recording run-time information and aids developers in debugging. However, effective logging requires a balance: logging too little can result in missing critical events, while excessive logging can cause important information to be overlooked, besides increasing storage and resources cost \cite{zhu2015learning}.

Since LLMs tend to generate verbose answers \cite{saito2023verbosity}, they may introduce more logs than a human would when prompted. This research question explores how the quantity and placement of logs compare between human-written and GPT-4o mini generated logs. Specifically, we want to determine if GPT-4o mini places logs in the same code blocks as humans.

\header{Approach.} To analyze how GPT-4o mini generates logs compared to our GT, we begin by extracting all log statements from each file in our dataset. For each log, we collect its ingredients: the log level, variables, text, and its location in the code. We determine location information using the AST blocks the log was introduced in, as described in Section \ref{sec:study_design}, and we call this the log's code path.

Next, we calculate the number of logs in each log path for both the GT and LLM files by summing the logs per path. We then merge these counts based on file name and log path, resulting in a dataset of the number of logs between GT and LLM per path for easier comparison. Using this merged dataset, we calculate the average of the following metrics: \textbf{Overlogging} refers to cases where the LLM inserted logs but the GT did not, while \textbf{Underlogging} captures the instances where the LLM failed to insert logs in places where the GT did.

To further analyze human and LLM logging practices, for the next two metrics, we filter our merged dataset to include only paths that contain at least one GT log. For these paths, we calculate \textbf{Coverage}, as a binary metric indicating whether an LLM-generated log appears in a position where a GT log exists, which is either 0 (Not Covered) or 1 (Covered), and \textbf{Quantified Coverage (QC)} as the ratio of LLM logs to GT logs in each path.

In summary, these are our four key metrics:

\begin{itemize}
    \item \textbf{Overlogging:} The proportion of paths where the LLM added logs but the GT did not. If a path contains zero GT logs and at least one LLM log, we mark that path as overlogged, and we report the average of these cases across the dataset.
    \item \textbf{Underlogging:} The proportion of paths where the GT added logs but the LLM did not. If a path contains one or more GT logs but the LLM had zero logs, we mark that path as underlogged, and we report the average of these cases across the dataset.
    \item \textbf{Coverage:} A binary indicator of whether the LLM inserted at least one log in a path where the GT logged. For each path with at least one GT log, if the LLM also includes one or more logs, we mark it as covered (1); otherwise, it is marked as not covered (0).
    \item \textbf{Quantified Coverage (QC):} The ratio of LLM logs to GT logs in each path where the GT logged. For example, if a path has 3 GT logs and 2 LLM logs, the QC is 2/3 since the LLM missed one log in that position. A QC value of 1 indicates an exact match in log count between the LLM and the GT in that path.
\end{itemize}

\header{Results.} \Cref{tab:rq1-results} presents our results based on log placement. GPT-4o mini achieved a Coverage rate of 63.91\%, which indicates a moderate level of similarity in adding logs in the same code paths as GT. This reflects that the LLM generally captures human behavior about which particular points in the code need to introduce a log. However, Coverage does not take into consideration the amount of logs added in those blocks. To address this, we examine the QC, which resulted in 68.03\%. This result shows that even in locations where the GT and LLM agree that a log is needed, the LLM tends to insert slightly more quantity of logs than human developers would. This reflects a tendency toward verbose logging at those instances.

This tendency becomes clearer when we examine the Overlogging rate of 88.66\%, which strongly confirms previous observations about LLM verbosity \cite{saito2023verbosity}. This indicates that GPT-4o mini frequently introduces logs in code blocks where human developers determined logging was unnecessary. Such behavior may reflect GPT-4o mini's preference to be on the side of caution, trying to log important information, but at the risk of adding excessive logs in production environments.

In contrast, the low Underlogging rate of 4.75\% suggests that GPT-4o mini is unlikely to miss logging positions that humans consider important. This difference between Overlogging and Underlogging rates suggests that the LLM prioritizes comprehensive coverage, rather than missing logging critical paths in the code.

\begin{table}
\centering
\caption{Coverage and verbosity metrics based on log placement across code paths}
\begin{tabular}{lrrrr}
\toprule
\textbf{Metric} & \textbf{Result} \\
\midrule
Coverage (binary)                   & 63.91\%  \\
Quantified Coverage (ratio)         & 68.03\%  \\
Overlogging             & 82.66\%  \\
Underlogging            & 4.75\%  \\
\bottomrule
\label{tab:rq1-results}
\end{tabular}
\end{table}

\begin{tcolorbox}[colback=green!5!white,colframe=green!75!black,title=Answer to RQ1]
Our results show that GPT4-o mini prioritizes comprehensive logging coverage, as shown by its low underlogging rate (4.75\%) and moderately high coverage (63.91\%). However, that results in generating a large number of logs, which is 5.15 times more than those written by humans. Even in cases where both humans and GPT-4o mini logged, the LLM tends to include more logs overall, with a rate of 68.03\%.
\end{tcolorbox}

\label{sec:rqii}
\subsection{RQ2: \rqii}

\header{Motivation.} In this research question, we aim to understand the quality of logs generated by GPT-4o mini in comparison to human-written logs. Specifically, we focus on instances where GPT-4o mini placed a log in the same position as the human developer. We evaluate GPT-4o mini performance based on three logging ingredients: levels, variables, and text. Understanding how well GPT-4o mini performs at generating these elements is crucial for evaluating its potential as a logging generator tool in software development, particularly in the context of ML applications where logging plays a critical role in monitoring model pipelines.

\header{Approach.} We evaluate log quality by first matching each LLM-generated log with its corresponding GT based on their paths (see Section \ref{sec:study_design}). To ensure a fair comparison, we calculate the evaluation metrics described below only for log pairs that share the same code path. Once matched, we assess the quality of the logging ingredients as follows:

\subsubsection*{1. Logging Levels}
Following prior work \cite{li2023exploring}, \cite{li2021deeplv}, we use two key metrics:

\begin{itemize}
    \item\textbf{Level Accuracy (L-ACC)} measures the percentage of log statements where GPT-4o mini correctly generates the exact same logging level as the GT.

    \item\textbf{Average Ordinal Distance (AOD)} quantifies how far off the predicted logging level is from the correct one. A higher AOD indicates closer alignment, even if the exact level was not predicted. We follow the Python logging library’s standard logging levels in ascending severity: \textit{debug}, \textit{info}, \textit{warning}, \textit{error}, and \textit{critical} \cite{logging_library}. The formula is:

\[
AOD = \frac{\sum_{i=1}^N (1-Dis(a_i, s_i)/MaxDis(a_i))}{N}
\]
where $N$ represents the total number of log statements across the dataset, while $Dis(a_i,s_i)$ denotes the distance between the actual logging level $a_i$ and the generated logging level $s_i$. $MaxDis(a_i)$ represents the maximum possible distance from the actual log level $a_i$ to the most distant level in the severity scale.
\end{itemize}

\subsubsection*{2. Logging Variables}
Each log may include zero or more variables. We define Variable Coverage as the proportion of GT variables correctly captured by GPT-4o mini. For example if the GT contains \textit{\{var1, var2, var3\}} and GPT-4o mini outputs \textit{\{var1, var3\}}, then Variable Coverage = 2/3.

\subsubsection*{3. Logging Text}
We evaluate the generated texts based on the following metrics used in Natural Language Processing (NLP) to evaluate automatically generated text: 

\begin{itemize}
    \item\textbf{BLEU} evaluates the quality of automatically generated text by measuring how many n-grams (short word sequences) it shares with the reference text \cite{papineni2002bleu}. A higher score indicates that more of the original phrasing is preserved. The score ranges from 0, meaning the texts are completely different, to 1, meaning they are identical. BLEU can be calculated using four different values of n-grams, such as BLEU-1 through BLEU-4.

    \item\textbf{ROUGE} is a set of metrics used to evaluate summarization tasks \cite{lin2004rouge}. We use the ROUGE-L variant, which is based on the Longest Common Subsequence (LCS). It identifies the longest sequence of words that appears in both texts in the same order, though not necessarily consecutively.  The values range from 0 to 1, with higher values indicating better similarity to the reference text.

    \item\textbf{METEOR} is a metric based on the harmonic mean of unigram precision and recall, with recall weighted higher than precision \cite{banerjee2005meteor}. METEOR considers synonyms, word stems, and sentence structure. It is designed to better align with human judgment and penalizes cases where the correct words are present but arranged incorrectly. The score ranges from 0 (completely different) to 1 (identical).

    \item\textbf{Levenshtein distance} is a metric for measuring the difference between two texts by calculating the minimum number of single-character edits required to change one text into the other \cite{levenshtein1966binary}. This metric provides an estimate of how much effort would be needed to modify the generated log so that it matches the GT. Following prior work \cite{mastropaolo2024leonid}, we utilize the normalized token-level Levenshtein distance (NTLev) between the generated log text and the GT. Normalization involves dividing the computed distance by the length of the text, resulting in a score between 0 (identical) and 1 (completely different).
\end{itemize}

\begin{table}
\centering
\caption{Evaluation metrics on log ingredients generated by GPT-4o mini}
\label{tab:rq2_results}
\begin{tabular}{lll}
\toprule
\textbf{Ingredient} & \textbf{Metric} & \textbf{Results} \\
\midrule
\multirow{2}{*}{Level} 
    & Level Accuracy (L-ACC)       & 59.19\%     \\ 
    & Average Ordinal Distance (AOD) & 84.34\%     \\ 
\midrule
Variables 
    & Variable Coverage       & 40.58\%      \\ 
\midrule
\multirow{8}{*}{Text} 
    & BLEU-1                       & 0.178      \\
    & BLEU-2                       & 0.095       \\
    & BLEU-4                       & 0.050       \\
    & METEOR                       & 0.328       \\
    & ROUGE-1                      & 0.341       \\
    & ROUGE-2                      & 0.121       \\
    & ROUGE-L                      & 0.316       \\
    & Levenshtein                  & 0.735      \\
\bottomrule
\end{tabular}
\end{table}

\header{Results.}
\Cref{tab:rq2_results} presents the performance of GPT-4o mini in generating each log ingredient at the file level.

\subsubsection*{1. Logging Levels}
GPT-4o mini exactly matches the level defined by developers in about 59\% of the cases across all the dataset. However, the AOD score of 84\% indicates that, even when the LLM fails to generate the exact level, it often selects a level close in severity.

\subsubsection*{2. Logging Variables}
GPT-4o mini achieves a Variable Coverage of 40\%, indicating that GPT-4o mini frequently misses the variables included in GT logs. This suggests that GPT-4o mini finds it difficult to identify the most relevant run-time data to include in the log, despite the file-level context giving access to a broader set of declared variables.

\subsubsection*{3. Logging Text}
GPT-4o mini's generated log texts demonstrate moderate alignment with the GT. The BLEU scores (BLEU-1: 0.178, BLEU-2: 0.095, BLEU-4: 0.050) suggest that the model produces texts with relatively few exact n-gram overlaps with the GT logs. However, a METEOR score of 0.328 and ROUGE scores (ROUGE-1: 0.341, ROUGE-2: 0.121, ROUGE-L: 0.316) indicate that the generated texts capture the general meaning of the GT logs. 

Overall, these results indicate that GPT-4o mini's generated log texts capture the meaning of the GT log texts (moderate ROUGE score), but it uses a different vocabulary or phrases the texts differently (low BLEU score). This is also supported by the high Levenshtein distance of 0.735, which reveals that substantial editing would be required to transform the GPT-4o mini generated texts into their human-written counterparts.

\begin{tcolorbox}[colback=green!5!white,colframe=green!75!black,title=Answer to RQ2]
GPT-4o mini shows moderate quality but significant limitations when generating logs compared to human developers. While it matches the log levels with 59.19\% exact accuracy and 84.34\% AOD, it identifies only 40.58\% of variables that human logged. Its log texts are similar in meaning to human logs (ROUGE-L: 0.316) but uses different vocabulary and phrasing (BLEU-4: 0.050), and require substantial editing to match human logs (Levenshtein: 0.735).
\end{tcolorbox}

\label{sec:rqiii}
\subsection{RQ3: \rqiii}

\header{Motivation.} Previous research has mostly evaluated the generation of logs at the function level, either by providing a specific line of code where to log \cite{li2023exploring} or by asking the LLM for the specific line where the log should be added \cite{xu2024unilog}. In contrast, our study focuses on evaluating LLMs in the task of generating logs for a complete file. In this research question, we aim to understand the specific challenges that arise when using an LLM to log complete files. Our goal is to identify the limitations and opportunities.

\header{Approach.}
We classify all the LLM-generated logs paired with human logs into four categories:

\begin{itemize}
    \item \textbf{Underlogging}: The LLM did not introduce a log in a path where the GT did.
    \item \textbf{Overlogging}: The LLM introduced a log in a path where the GT did not.
    \item \textbf{Different Level}: Both LLM and GT introduced a log in the same position, but used different log levels.
    \item \textbf{Different Variables}: Both LLM and GT introduced a log in the same position, but logged different variables. For this category, we consider pairs of logs where both GT and LLM have at least one variable.
\end{itemize}

Using a 95\% confidence level and a 5\% margin of error, we obtain a stratified sample of 384 paired logs for manual analysis. \Cref{tab:rq3-dataset} shows the distribution of logs within the stratified sample. For each pair of logs, we manually analyze the GT and LLM files where the log was introduced. This analysis involves examining the surrounding code to better understand the context in which the log statement was introduced. In case of log level, we also analyze the repository's logging level preferences based on the quantity of different levels of severity used in the repository.

\begin{table}
\centering
\caption{Categories of challenges in LLM-generated logs}
\begin{tabular}{lrrrr}
\toprule
   & \textbf{\# Paired logs} & \textbf{\# Sampled logs} & \textbf{Proportion}\\
\midrule
Underlogging        & 5,382      & 18  & 4.7\% \\
Overlogging         & 98,208     & 328 & 85.9\% \\
Different Level     & 6,018     & 20  & 5.2\% \\
Different Variables & 4,914     & 16  & 4.2\% \\
\midrule
Total               & 114,522   & 382 & 100\% \\
\bottomrule
\label{tab:rq3-dataset}
\end{tabular}
\end{table}

\header{Results.}
We examine each category of logging differences between GPT-4o mini and GT below.

\textbf{Overlogging} was the most frequent issue found in our dataset (85.8\%). After manual analysis, we found that most of these instances reflected that the LLM often inserts logs at the start or end of functions (179/328) or blocks (120/328). In most examples, the LLM interpreted the beginning of a function or a block as an important process that was about to begin, even when the function was not implemented, as shown in \Cref{lst:overlogging-notimplemented}.

\begin{center}
\begin{minipage}{0.43\textwidth}
\begin{lstlisting}[caption={An example of overlogging: Not implemented function with introduced log}, label={lst:overlogging-notimplemented}]
def sync_optimizer(self, optimizer: Optimizer):
    logger.info("Syncing optimizer across processes.")
    pass
\end{lstlisting}
\end{minipage}
\end{center}

Moreover, when logging about the start or completion of a process, the LLM tends to include variables (151/328). However, it also frequently omits them (112/328). For example, in \Cref{lst:overlogging-notimplemented} the log message references the optimizer process, but fails to include details about the optimizer object itself despite it being a function parameter. The LLM exhibited a pattern of adding logs immediately before raising exceptions (40/328), as demonstrated in line 5 of \Cref{lst:overlogging-example2}. Line 2 of \cref{lst:overlogging-example2} shows an example of the type of variables the LLM would include in logs, which are generally variables part of that block, and it can infer variables that are part of an object if they were used later in the code.

\begin{center}
\begin{minipage}{0.43\textwidth}
\begin{lstlisting}[caption={An example of overlogging: Line 2 shows a log added at the start of a function, while log 5 represents a log introduced just before raising an exception}, label={lst:overlogging-example2}]
def _intersect(self, dataloader):
    logger.debug("Calculating intersection with dataloader of type: %s", type(dataloader.backend))
    if not isinstance(dataloader.backend, NoBackend):
        msg = 'Intersection can only be calculated between same backends (NoBackend), instead get {}'.format(type(dataloader.backend))
        logger.error(msg)
        raise Exception(msg)
\end{lstlisting}
\end{minipage}
\end{center}

\textbf{Underlogging} occurred less often (4.7\%), but typically involved missing variable changes in long blocks of code (5/18), or failing to log process state within conditions (10/18). While nearly half of the overlogging cases were logs placed at the beginning of code blocks, in underlogging, we observed instances where GPT-4o mini failed to insert logs within long code blocks (i.e., a long main function) even when introducing new inner blocks (i.e., \textit{for} loop). This pattern is demonstrated in \cref{lst:underlogging}, where the GT logs before evaluation on line 112; in contrast, the LLM does not log until the ending of the main function. This suggests that GPT-4o mini has trouble understanding what is important in large chunks of code. Also, the LLM seems to infer what the code is doing by heavily relying on the function name where the log is added.

\begin{center}
\begin{minipage}{0.43\textwidth}
\begin{lstlisting}[caption={An example of underlogging: The LLM miss logging in long blocks of code}, label={lst:underlogging}]
def main():
    # 153 lines before this one
    # GT log
    logger.info("Evaluate the following checkpoints: %s", checkpoints)
    for checkpoint in checkpoints:
        global_step = checkpoint.split('-')[-1] if len(checkpoints) > 1 else ''
        model = load_model(args, checkpoint=checkpoint)
        result = evaluate(args, model, tokenizer, prefix=global_step)
        result = dict(((k + ('_{}'.format(global_step) if global_step else ''), v) for (k, v) in result.items()))
        results.update(result)
    # LLM log
    logging.info('Main function completed.')
    return results
\end{lstlisting}
\end{minipage}
\end{center}

\textbf{Different levels} (5.3\%) frequently involved level mismatches between \textit{info} and \textit{debug} (8/20). In cases where the mismatch involved other levels, such as \textit{debug} vs. \textit{warning}, the log level chosen by the LLM (5/20) or by the developer (2/20) was more appropriate based on the severity level defined by the Python logging library \cite{logging_library}. A key insight was that in the majority of level mismatches (13/20), the developers chose levels that matched the most commonly used level in that particular project, suggesting that project-specific logging preferences influence log level selection. \Cref{lst:different-level} shows an example of logs where GT and LLM exhibit the same logging intent, but use different levels.

\begin{center}
\begin{minipage}{0.43\textwidth}
\begin{lstlisting}[caption={An example of logs with different level: Log with the same intent but different log level}, label={lst:different-level}]
for i, point in enumerate(datapoints):
    if i % log_every_n == 0:
        # GT log
        logger.info("Featurizing datapoint %i" % i)
        # LLM log
        logger.debug(f'Featurizing datapoint {i}/{len(datapoints)}.') 
        try:
            features.append(self._featurize(point, **kwargs))
        except:
            features.append(np.array([]))
\end{lstlisting}
\end{minipage}
\end{center}

\textbf{Different variables} (4.3\%) often reflected different log intents (7/18), suggesting that different log purposes led to different logged variables. GPT-4o mini captured some but not all of the variables that developers did (5/16), indicating a partial understanding of which variables are important in each code block. In a few cases, the LLM missed all the variables that developers considered important to log (3/18).

One of the main challenges is that GPT-4o mini seems not to have enough context of external class definitions; therefore, it is unable to suggest relevant variables external to the file. \Cref{lst:different-vars} shows an instance where the LLM lacked context about the objects defined in imported classes.

\begin{center}
\begin{minipage}{0.43\textwidth}
\begin{lstlisting}[caption={An example of logs with different variables: LLM does not have context of Topology's object variables}, label={lst:different-vars}]
def _debug_dump_topology(topology: Topology, resource_manager: ResourceManager) -> None:
    for i, (op, state) in enumerate(topology.items()):
        # GT log
        logger.debug(
            f"{i}: {state.summary_str(resource_manager)}, "
            f"Blocks Outputted: {state.num_completed_tasks}/{op.num_outputs_total()}"
        )
        # LLM log
        logger.debug(f'Debug dump for operator {op}: {state}')
\end{lstlisting}
\end{minipage}
\end{center}

\begin{tcolorbox}[colback=green!5!white,colframe=green!75!black,title=Answer to RQ3]
Our manual analysis of GPT-4o mini generated logs shows that the main challenge is overlogging (85.8\%), with most logs introduced at the start or end of a function. Second, the LLM underlogs (4.7\%), particularly in large code blocks. Third, in cases with different log level (5.3\%) the LLM does not align with project-specific logging conventions. Finally, in case of different variables captured between humans and GPT-4o mini (4.3\%), the variables introduced by the LLM sometimes miss important variables that are present in the GT, particularly when those variables were from imported classes outside the file's context.
\end{tcolorbox}

\section{Implications}
\label{sec:implications}
\header{Review LLM output to reduce overlogging.} Our findings highlight that while LLMs can generate logging statements that align with general developer intentions, they often overcompensate by adding logs in places that emphasize process flow rather than critical state or variable changes. This behavior suggests that LLMs may clutter the code with excessive logs, which reduces the usefulness of logging for debugging and monitoring. Developers should review and filter LLM-generated logs rather than blindly adopt them in their code.

\header{Add context related to imported classes or functions.}
To address the problem of logging better variables, researchers could look into adding context about external methods or classes related to the file. This additional context may help the LLM generate relevant variables. Li et al. \cite{li2024go} explored this idea in the problem of generating logs for a function, but it could be expanded to generate logs for a file.

\header{Incorporate repository-specific logging configurations.} We observed that some repositories define custom logging levels or adjust the severity of existing ones. These repository-specific configurations are not typically captured in research related to log generation. To enable full automation of logging files, future work should integrate these custom configurations into the prompting context.

\header{Refine prompt with pipeline step context.} Using LLMs for prompting ML applications may benefit from introducing relevant examples of logs related to specific pipeline steps. This could improve the performance of generated logs, since ML applications involve critical processes such as model training or data preprocessing pipelines.

\header{File-level benchmarks for automated logging.} In this paper we explored evaluating GPT-4o mini in the file-level context. We focus on logs introduced at certain paths and compare them. Future benchmarks could go further and focus on assessing whether these introduced logs actually support developers in tasks like debugging or root cause analysis.

\section{Limitations}
\label{sec:limitations}
In this section, we discuss threats to the validity of our study.

\header{Internal Validity.} 
The internal validity of our study is affected by three main factors. First, we use GPT-4o mini for our study, which is pre-trained with information up until October 2023, while our data was collected from GitHub in October 2024. As a result, there is a potential risk of data leakage for repositories that remained unchanged during that period. However, we note that the outputs generated by the LLM are not closely aligned with the GT. This suggests that, even if the LLM had seen parts of the repositories during training, it does not appear to result in memorization or reproduction of the original logs. Therefore, while the theoretical risk of data leakage exists, its practical impact on our findings is likely minimal. Second, the effects of using different prompts change the outcome of the results. To minimize this threat, we crafted a prompt that explicitly asks for log quality instructions related to logging placement and logging ingredients. Third, we made prompt design choices to manage the token limitations of GPT-4o mini that may have affected result quality. To prevent exceeding GPT-4o mini's context window, we removed documentation and existing log comments to avoid token cutoff, while still providing the complete source file to maintain code context. Previous studies \cite{li2023exploring} have found that prompting LLMs and removing documentation actually decreases the quality of the logging ingredients. In contrast, the same study found that prompting the complete file for logging helps the LLMs get more context about the functions in the code, which translates to better logging quality output from the LLM. Future work should focus on refining prompts to include documentation while incorporating file context, and limiting the prompt to output only the necessary log statements to address the LLM output token limitations.

\header{External Validity.} 
First, we used GPT-4o mini because, at the time of our study, it was a state-of-the-art model with a large context window. While other LLMs may show different performance, we believe that the challenges identified in the paper would remain important, albeit with different frequencies. Second, we focused our research on open-source projects hosted on GitHub, which may not fully represent the wider spectrum of machine learning projects, especially proprietary ones. To mitigate this, we utilized established best practices in mining software repositories, which are selecting active, popular, and mature ML projects \cite{munaiah2017curating}. Furthermore, our dataset includes a diverse range of ML projects, which also have been studied before in the literature. Second, our study was made for Python repositories, which may affect the generalizability of our findings to other languages. However, previous research found that Python is the most utilized language for machine learning applications \cite{gonzalez2020ml_universe}. Future work can explore how our findings extend to other languages. Third, we focus on generating logs using the \textit{logger} library, and did not explore generating logs for ML-specific logging libraries such as MLFlow or Weights \& Biases (Wandb). However, in our dataset of 4,073 files, only 32 files used Wandb and 118 used MLFlow. Future work could evaluate the quality of LLM-generated logs for ML-specific logging libraries using a larger dataset.

\section{Related Work}
\label{sec:related_work}
\subsection{Empirical studies on logging practices}

Yuan et al. \cite{yuan2012characterizing} conducted an empirical study on logging practices in open-source systems written in C/C++. Their analysis revealed that developers extensively utilize log statements and regularly modify them to enhance both debugging capabilities and system maintenance processes. Recently, Foalem et al. \cite{foalem2024logging_ml} examined logging practices in ML applications. They found that ML-based applications use both general logging libraries and ML specific libraries. They also found that the most used logging levels in ML applications are \textit{info} and \textit{warning}.

Fu et al. \cite{fu2014developers} analyzed logging practices across two Microsoft industrial projects, specifically examining which types of code blocks developers typically choose to log. Their analysis showed that developers log unexpected situations where the system would throw an error, and they also log critical execution points (i.e., execution paths). Kabinna et al. \cite{kabinna2018examining} found that 20–45\% of introduced logging statements are changed at least once in the project's lifetime.

These foundational studies establish our understanding of logging practices, including both traditional and ML applications. Our research complements this work by exploring how GPT-4o mini generates log statements at the file level for ML files and identifying its limitations.

\subsection{Automating logging statements}
Studies in logging statement automation are divided into addressing the problems of  \textit{what-to-log} and \textit{where-to-log}. \textit{What-to-log} studies are generally divided into three subtasks: deciding the appropriate log level \cite{li2021deeplv, li2017log, liu2022tell}, selecting log variables \cite{liu2019variables, dai2022reval, yuan2012improving}, and generating a log message \cite{ding2022logentext}. \textit{Where-to-log} studies focus on suggesting the best log placement \cite{li2020shall, zhao2017log20}. 

Recently, there have been studies that tackle the problem of generating complete logs by deciding the level, variables, message, and where it should be introduced. Mastropaolo et al. \cite{mastropaolo2022lance} introduced LANCE, which is a tool that automatically generates a proper log level, message, variables, and location. Mastropaolo et al. \cite{mastropaolo2024leonid} then improved on their approach with LEONID, which also generated a complete log as LANCE, but it also has the capability of deciding if a method needs a log or not. Xu et al. \cite{xu2024unilog} proposed a warmup and in-context learning (ICL) approach to enhance log generation. Since this is a prompt approach, it can be used with any LLM as a backbone. Xie et al. proposed FastLog \cite{xie2024fastlog}, which uses a token classification model to locate where a log statement should be added, and then uses a Seq2Seq model to generate a complete logging statement for that position. Li et al. proposed SCLogger \cite{li2024go}, which introduces domain knowledge to extend the context of the prompt to include logging-related context. Tan et al. introduced AL-Bench \cite{tan2025bench}, which includes a dynamic evaluation of logs at run-time.

While existing tools target function-level log generation for general applications, our work complements existing work by examining file-level log generation in ML projects using GPT-4o mini. This larger context introduces unique challenges, such as overlogging, adapting to project-specific logging conventions, and managing variables available in external classes.

\section{Conclusion}
\label{sec:conclusion}
Our findings highlight both the potential and limitations of GPT-4o mini for file-level automated log generation in ML applications. The LLM showed moderate effectiveness in identifying where to place logs. At the same time, this high coverage comes at the cost of generating 5.15 times more log statements than human developers, which could limit its practical usefulness. In terms of evaluating logging ingredients, the LLM achieves a 59.19\% exact log match. However, GPT-4o mini for file-level logging struggles with selecting relevant variables, covering only 40.58\% of variables in the GT. Log texts are similar in meaning to human logs but use different vocabulary and phrasing. Our manual analysis further revealed particular issues of logging complete files, such as logging at the start or end of functions, skipping the introduction of logs in large code blocks, and the lack of adherence to project-specific logging preferences.

Overall, our results indicate areas of improvement for file-level automated logging. We suggest adding relevant context related to each file in the prompt to generate better logs, as well as to extract some samples related to the ML pipeline to improve significant variables or messages specific to ML logs.

\bibliographystyle{IEEEtran}
\bibliography{references}

\end{document}